\documentclass[a4paper]{jpconf}
\usepackage{graphicx}
\begin{document}
\title{The Omega Effect for Neutral Mesons}

\author{Sarben Sarkar}

\address{Department of Physics, King's College London, Strand, London WC2R~2LS, UK}

\ead{sarben.sarkar@kcl.ac.uk}

\begin{abstract}
The possible role of decoherence due to space-time foam is discussed within
the context of two models, one based on string/brane theory. and the other
based on properties of black hole horizons in general
relativity. It is argued that the density matrix satisfies a dissipative master
equation, primarily from the study of renormalization group flows \ in
non-critical string theory.This interpretation of the zero mode of the Liouville
field as time leads necessarily to the CPT operator being ill defined. One
striking consequence is that the quantum mechanical correlations of pair
states of neutral mesons produced in meson factories are changed from the
usual EPR state. The magnitude of this departure from EPR correlations is
characterised by a parameter $\omega$. The predicted value of $\omega$ is very small or zero. However
it is shown explicitly that the the non-vanishing of $\omega$ is only a feature of the model based on
string/brane theory.
\end{abstract}

\section{Introduction}
The important feature of General Relativity (GR), the classical field theory
describing gravity, is the fact that space-time is not simply a frame of
coordinates, on which events take place, but is itself a dynamical entity. The
quantisation of the theory, poses a problem, given that the coordinates of
space-time themselves appear ``fuzzy''. The fuzziness of space-time is
associated with microscopic quantum fluctuations of the metric field, which
may be singular. For instance, one may have Planck size ($10^{-35}$ m) black
holes, emerging from the quantum gravity ``vacuum'', which may give space-time
a ``foamy'', topologically non-trivial structure. The semi-classical study of
black holes in GR has led to the prediction of Hawking radiation. In classical
GR simple solutions for black holes exhibit event horizons. This was
interpreted quantum mechanically in terms of lack of communication between
Hilbert spaces within and outside the horizon. This is at the basis of the
information paradox\cite{HawkingPage}. Simple arguments relating to tracing
over states then lead to a description in terms of non-pure density matrices.
These observations have led to a vigorous debate concerning the validity of
unitary evolution for quantum gravity. Initially S. Hawking believed that
there was information loss; he and others have recently
claimed~\cite{hawkingrecent} that information is not lost in the presence of a
black hole, but rather it is entangled in a holographic way with the portion
of space-time outside the horizon. There seems currently more and more adherents
to this point of view. However in my opinion the situation is not resolved. This is
primarily because we do not have a theory of quantum
gravity. Consequently there are always certain loose ends in the arguments
that are proposed. As an example the original argument of Hawking has been
criticized~\cite{hayward} on the grounds that horizons are mathematical
constructs and are not as physically relevant as trapping surfaces. It is then
argued that evaporating black holes can evade the information paradox owing to
the presence of an inner and outer trapping surface. The recent arguments of
Hawking and other (string) theorists are somewhat special in their details and
require constructions such as the extremality of black holes, anti-de Sitter
(supersymmetric) space-times and the Euclidean formulation for summing over
space-time geometries. Such considerations have led to a point of view
labelled as holography and black hole complementarity\cite{giddings}%
,\cite{thooft} .The special features in these arguments make it hard to
make general statements about unitarity or the lack of it in quantum evolution.

We will take a somewhat different point of view which will put us firmly in
the camp of those who believe that there is non-unitary evolution due to
quantum gravity. Much of the discussion using string theory is based on
critical strings where the demands of world sheet conformal and target space
Lorentz invariance constrain the allowed values of space-time dimension (see
e.g. \cite{zwiebach}). Consequently changes in space-time backgrounds cannot
be accommodated. Surely a proper theory of quantum gravity needs to be able to
handle this. This is a fundamental problem. An approach which takes some steps
to address these issues is based on the theory of non-critical
strings\cite{EllisMavNanop}  also referred to as two dimensional
gravity coupled to matter. The restoration of conformal invariance for a
number of spatial dimensions different from the critical value is addressed
through the introduction of the Liouville field. Furthermore by studying the
recoil of branes and the induced backreaction on space-time, it possible to
consider some situations where the space-time metric and hence geometry is
changed. When $p$ space dimensional Dirichlet (Dp) branes are considered, it
is no longer necessary to concentrate on the properties of microscopic black
holes and the inherent issues of trapped surfaces. In particular Dp branes
with $p=0$ exist in some string theories such as bosonic, type IIA and type I.
Furthermore even when elementary D particles cannot exist consistently there
can be effective D-particles formed by the compactification of higher
dimensional D branes. Moreover D particles are
non-perturbative constructions since their masses are inversely proportional
to the the string coupling ${g_{s}}$ . The introduction of the Liouville field
opens up also an interesting possibility. It was suggested by
Shore~\cite{shore} that a renormalization group scale which was local on the
world sheet was a useful way of analyzing string theory sigma models (in the
sense of the Zamolodchikov C-function) and subsequently~\cite{EllisMavNanop}
this scale was identified with the Liouville field. Furthermore the zero mode
of the Liouville field~ can be identified with time on the basis of transformation
properties of vertex operators associated with D particles \cite{mavromatos}.
The Zamolodchikov C theorem (for general two dimensional field theories)
requires that the C-function (a function in the space of all couplings,
i.e. theory space) is a non-negative and non-increasing function
on a renormalization group trajectory. The renormalization group trajectory
can be identified with temporal evolution as just intimated and this leads,
because of the increase of entropy along the trajectories, to a non-unitary
evolution for the density matrix $\rho$. It will suffice for us to note that
the evolution has the form%

\begin{equation}
\frac{{\partial\rho}}{{\partial t}}=i\left[  {\rho,H}\right]  +\delta\not H
\rho\, \label{liouvqm}%
\end{equation}
where $H$ is the hamiltonian and $\delta\not H  \rho$ is formally written as%

\begin{equation}
\delta\not H  \rho={\beta^{i}}{G_{ij}}\frac{{\partial\rho}}{{\partial{p_{j}}}%
}. \label{nonunitary}%
\end{equation}
Here $G_{ij}$ is the so called Zamolodchikov metric, ${\beta^{i}}$ is the
renormalization group $\beta$ function for the couplings $g_{i}$
associated with vertex functions $p_{j}$ spanning theory space. Although the
identification of the Liouville field with the local renormalization group
scale is not rigorous, it is one important feature which influences our
joining of the "non-unitary" camp. The other important reason for us is
D-particle recoil which will be addressed in a later section.

The second model to be considered by us is concerned with the effect of
horizons and the consequent absence of unitarity but the formulation is not
supported by a formal theory like string theory. A different effective theory
of space-time foam has been proposed by Garay \cite{Garay}. The fuzziness of
space-time \ at the Planck scale is described by a non-fluctuating background
which is supplemented by non-local interactions. The latter reflects the fact
that at Planck scales space-time points lose their meaning and so these
fluctuations present themselves in the non-fluctuating coarse grained
background as non-local interactions. These non-local interactions are then
rephrased as a quantum thermal bath with a Planckian temperature. The quantum
entanglement of the gravitational bath and the two meson (entangled) state is
explicit in this model. Consequently issues of \textit{back reaction} can be
readily examined. Since the evolution resulting from the standard Lindblad
formulation does not lead to the $\omega$ effect, this manifestation of CPTV
is not the result of some arbitrary non-unitary evolution. Hence it is
interesting to study the above two quite distinct models (one motivated by
string theory and the other by the properties of black holes in general relativity) for clues concerning the
appearance of (and an estimate for the order of magnitude) of $\omega$.

Recently these fundamental issues have been brought into sharp focus through a
phenomenological analysis involving correlated neutral meson pairs. For
equations such as (\ref{liouvqm}), it is well known that initial pure states
evolve into mixed ones and so the S-matrix $\not S  $ relating initial and
final density matrices does not factorise, i.e.
\begin{equation}
\not S  \neq SS^{\dag} \label{Nonfactorisability}%
\end{equation}
where $S=e^{iHt}$. This is another way \ of saying that we have \ non-unitary
evolution. In these circumstances Wald~\cite{wald} has shown that CPT is
violated, at least in its strong form, i.e. there is \textit{no} unitary
invertible operator $\Theta$ such that
\begin{equation}
\Theta\overline{\rho}_{\mathrm{in}}=\rho_{\mathrm{out}}\mathrm{.}%
\end{equation}
This result is due to the entanglement of the gravitational fluctuations with
the matter system.

It was pointed out in \cite{Bernabeu}, that if the CPT operator is not well
defined then this has implications for the symmetry structure of the initial
entangled state of two neutral mesons in meson factories. Indeed, if CPT can
be defined as a quantum mechanical operator, then the decay of a (generic)
meson with quantum numbers $J^{PC}=1^{--}$ \cite{Lipkin}, leads to a pair
state of neutral mesons $\left\vert i\right\rangle $ having the form of the
entangled state%
\begin{equation}
\left\vert i\right\rangle =\frac{1}{\sqrt{2}}\left(  \left\vert \overline
{M_{0}}\left(  \overrightarrow{k}\right)  \right\rangle \left\vert
M_{0}\left(  -\overrightarrow{k}\right)  \right\rangle -\left\vert
M_{0}\left(  \overrightarrow{k}\right)  \right\rangle \left\vert
\overline{M_{0}}\left(  -\overrightarrow{k}\right)  \right\rangle \right)  .\label{CPTV}
\end{equation}
This state has the Bose symmetry associated with particle-antiparticle
indistinguishability $C\mathcal{P}=+$, where $C$ is the charge conjugation and
$\mathcal{P}$ is the permutation operation. If, however, CPT is not a good
symmetry (i.e. ill-defined due to space-time foam), then $M_{0}$ and
$\overline{M_{0}}$ may not be identified and the requirement of $C\mathcal{P}%
=+$ is relaxed~\cite{Bernabeu}. Consequently, in a perturbative framework, the
state of the meson pair can be parametrised to have the following form:
\[
\left\vert i\right\rangle    =\frac{1}{\sqrt{2}}\left(  \left\vert
\overline{M_{0}}\left(  \overrightarrow{k}\right)  \right\rangle \left\vert
M_{0}\left(  -\overrightarrow{k}\right)  \right\rangle -\left\vert
M_{0}\left(  \overrightarrow{k}\right)  \right\rangle \left\vert
\overline{M_{0}}\left(  -\overrightarrow{k}\right)  \right\rangle \right)
\\
  +{\frac{\omega}{\sqrt{2}}}\left| {\Delta \left( {\vec k} \right)} \right\rangle \label{omegaterm}
\]
where
\[\left| {\Delta \left( {\vec k} \right)} \right\rangle  \equiv \left| {{{\overline M }_0}\left( {\vec k} \right)} \right\rangle \left| {{M_0}\left( { - \vec k} \right)} \right\rangle  + \left| {{M_0}\left( {\vec k} \right)} \right\rangle \left| {{{\overline M }_0}\left( { - \vec k} \right)} \right\rangle \]
and $\omega=\left\vert \omega\right\vert e^{i\Omega}$ is a complex CPT
violating (CPTV) parameter. For definiteness in what follows we shall term
this quantum-gravity effect in the initial state as the \textquotedblleft%
$\omega$-effect\textquotedblright\cite{Bernabeu}. There is actually another dynamical
\textquotedblleft$\omega$-effect\textquotedblright\ which is generated during
the time evolution of the meson pair but this will not be discussed here \cite{bernabeu}.

The structure of the article will be the following: we commence our analysis
by discussing the omega effect.We will then proceed to discuss a string based
model for the omega effect. The difficulty of generating the omega effect
using other plausible approaches incorporating open systems and non-unitary
master equations is illustrated through a class of models which we dub the
thermal bath model.

\section{D-particles\label{sec:2}}
\begin{figure}[t]
\centering\includegraphics[width=7.5cm]{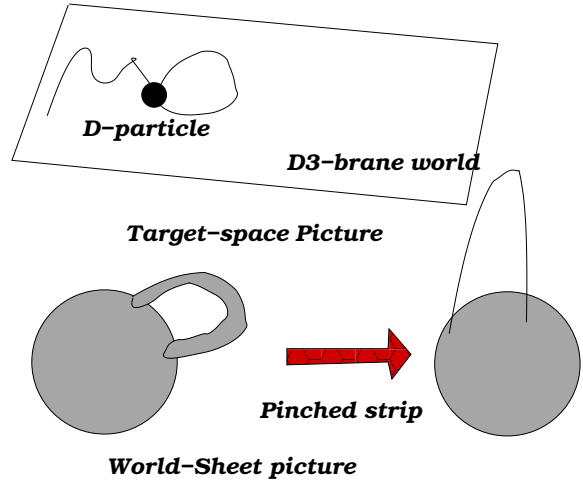}\caption{\emph{Upper
picture:} A Fluctuating D-particle is described by open strings attached to
it. As a result of conservation of fluxes~\cite{polch,polch2,johnson} that
accompany the D-branes, an isolated D-particle cannot occur, but it has to be
connected to a D-brane world through flux strings. \emph{Lower picture}:
World-sheet diagrams with annulus topologies, describing the fluctuations of
D-particles as a result of the open string states ending on them. Conformal
invariance implies that pinched surfaces, with infinitely long thin strips,
have to be taken into account. In bosonic string theory, such surfaces can be
resummed~\cite{szabo}. }%
\label{fig:dbranes}%
\end{figure}

\begin{figure}[th]
\centering\includegraphics[width=7.5cm]{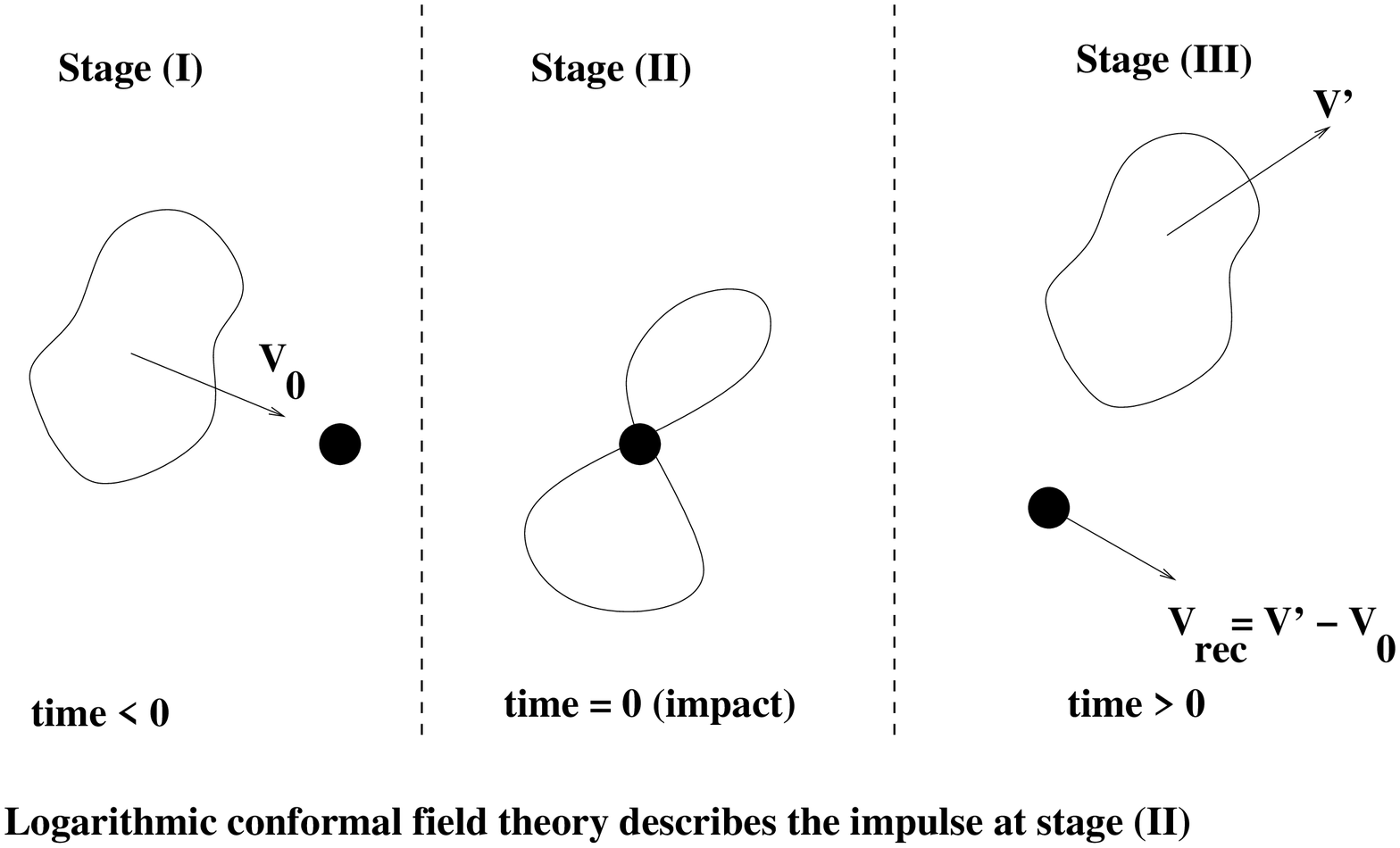} \hfill
\includegraphics[width=7.5cm]{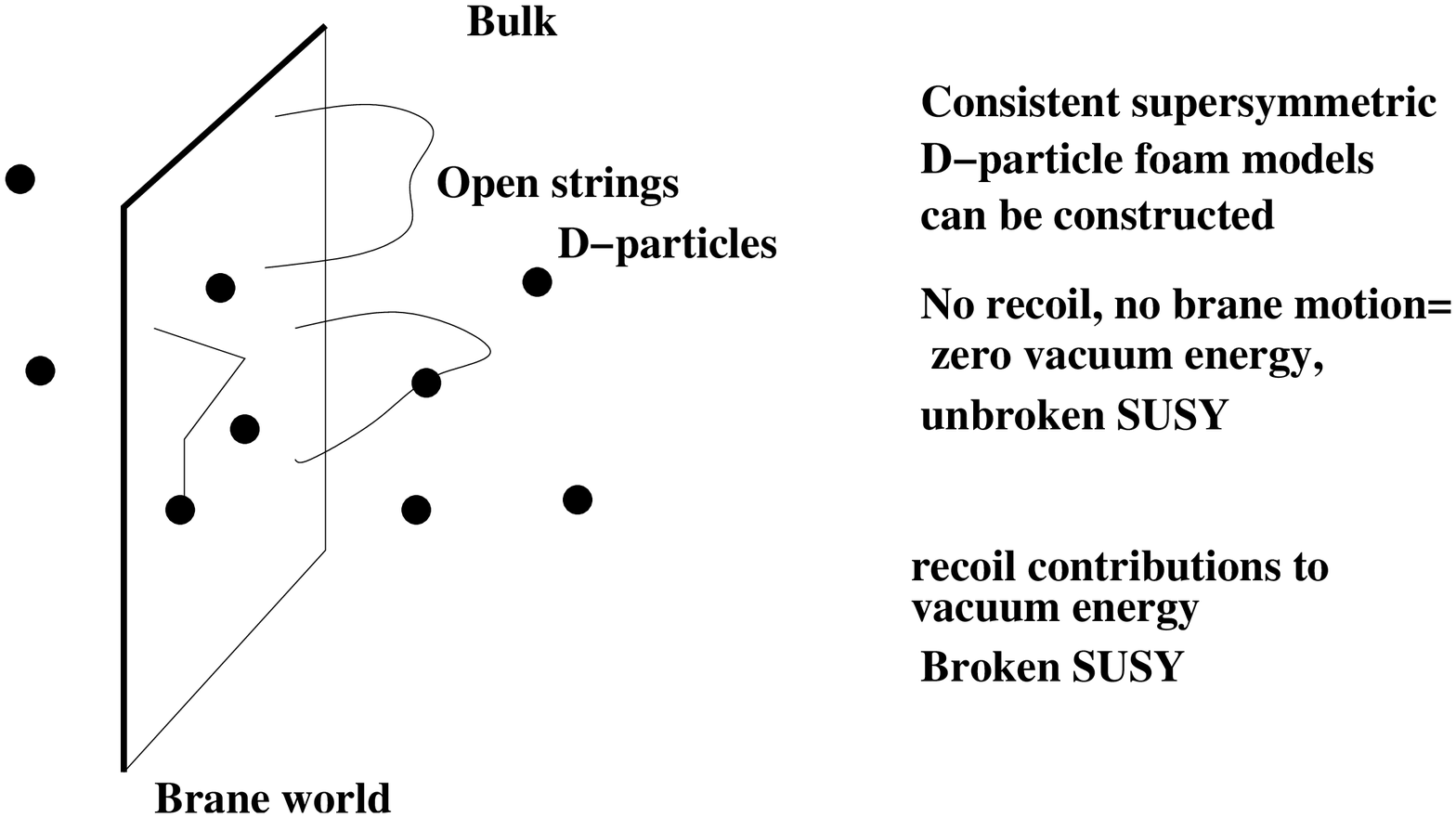} \caption{Schematic
representation of a D-foam. The figure indicates also the capture/recoil
process of a string state by a D-particle defect for closed (left) and open
(right) string states, in the presence of D-brane world. The presence of a
D-brane is essential due to gauge flux conservation, since an isolated
D-particle cannot exist. The intermediate composite state at $t=0$, which has
a life time within the stringy uncertainty time interval $\delta t$, of the
order of the string length, and is described by world-sheet logarithmic
conformal field theory, is responsible for the distortion of the surrounding
space time during the scattering, and subsequently leads to induced metrics
depending on both coordinates and momenta of the string state. This results on
modified dispersion relations for the open string propagation in such a
situation~\cite{Dfoam}, leading to non-trivial \textquotedblleft
optics\textquotedblright\ for this space time.}%
\label{fig:recoil}%
\end{figure}

String theory was found, in the so called first revolution, to contain quantum
gravity while in the second revolution the proposal of D(irichlet) brane
solitons~\cite{polch,polch2}, together with dualities brought about
relationships between the different string theories that had been proposed
earlier. In particular zero dimensional D-branes \cite{johnson} occur (in
bosonic and some supersymmetric string theories) and are known as
D-particles. The spectrum of open strings attached to a Dp brane (with $p>0$ )
contains a Maxwell field and the ends of the open string carry charge. The
associated Maxwell field is confined to the world volume of the D brane. Hence
a conventionally charged string cannot end on a D-particle. We will thus
restrict our consideration to neutral particles that are \textquotedblright
captured\textquotedblright\ \ by D-particles. The D-particle solitons are fundamental
scalar particles in the dual theories. Interactions in string theory are, as
yet, not treated as systematically as in ordinary quantum field theory where a
second quantised formalism is defined. The latter leads in a systematic way to
the standard formulations by Schwinger and Feynman of perturbation series. When
we consider stringy matter interacting with other matter or D-particles, the
world lines traced out by point particles are replaced by two-dimensional
world sheets. World sheets are the parameter space of the first quantised
operators ( fermionic or bosonic) representing strings. In this way the first
quantised string is represented by actually a two dimensional (world-sheet)
quantum field theory. An important symmetry of this first quantised string
theory is conformal invariance and the requirement of the latter does
determine the space-time dimension and/or structure. \ This symmetry leads to
a scaling of the metric and permits the representation of interactions through
the construction of measures on inequivalent Riemann surfaces \cite{strings}.
In and out states of stringy matter are represented by punctures at the
boundaries. The D-particles as solitonic states~\cite{polch2} in string theory
do fluctuate themselves, and this is described by stringy excitations,
corresponding to open strings with their ends attached to the D-particles. In the
first quantised (world-sheet) language, such fluctuations are also described
by Riemann surfaces of higher topology with appropriate Dirichlet boundary
conditions (c.f. fig.~\ref{fig:dbranes}). The plethora of Feynman diagrams in
higher order quantum field theory is replaced by a small set of world sheet
diagrams classified by moduli which need to be summed or integrated over
\cite{zwiebach}. In order to understand possible consequences for CPT due to
space-time foam we will have to characterise the latter. The model we will
consider is based on D-particles populating a bulk geometry between parallel
D-brane worlds. The model is termed D-foam~\cite{Dfoam} (c.f. figure
\ref{fig:recoil}), since our world is modelled as a three-brane moving in the
bulk geometry. As a result, D-particles cross the brane world, and thereby
appear as foamy flashing on and off structures for an observer on the
brane. 

Even at low energies $E$, such a foam may have observable consequences e.g.
decoherence \ effects which may be of magnitude $O\left(  \left[  \frac
{E}{M_{P}}\right]  ^{n}\right)  $ with $n=1,2$ where $M_{P}$ is the Planck
mass or change in the usual Lorentz invariant dispersion relations.The study
of D-brane dynamics has been made possible by Polchinski's realisation that
such solitonic string backgrounds can be described in a conformally invariant
way in terms of world sheets with boundaries \cite{polch2}. On these
boundaries Dirichlet boundary conditions for the collective target-space
coordinates of the soliton are imposed \cite{coll}. Heuristically, when low
energy matter given by a closed string propagating in a $\left(  d+1\right)
$-dimensional space-time collides with a very massive D-particle (0-brane)
embedded in this space-time, the D-particle recoils as a result \cite{kogan}.
We shall consider the simple case of bosonic stringy matter coupling to
D-particles and so we can only discuss matters of principle and ignore issues
of stability. However we should note that an open string model needs to
incorporate for completeness, higher dimensional D-branes such as the D3
brane. This is due to the vectorial charge carried by the string owing to the
Kalb-Ramond field. Higher dimensional D-branes (unlike D-particles) can carry
the charge from the endpoints of open strings that are attached to them. For a
closed bosonic string model the inclusion of such D-branes is not imperative
(see figure \ref{fig:recoil}). The higher dimensional branes are not pertinent
to our analysis however. The current state of phenomenolgical modelling of the
interactions of D-particle foam with stringy matter will be briefly summarised
now. Since there are no rigid bodies in general relativity the recoil
fluctuations of the brane and their effective stochastic back-reaction on
space-time cannot be neglected. As we will discuss, D-particle recoil in the
''tree approximation'' i.e. in lowest order in the string coupling $g_{s}$,
corresponds to the punctured disc or Riemann sphere approximation in open or
closed string theory, induces a non-trivial space-time metric. For closed
strings colliding with a heavy (non-relativistic) D-particle the metric has
the form \cite{mav2}
\begin{equation}
g_{ij}=\delta_{ij},\,g_{00}=-1,g_{0i}=\varepsilon\left(  \varepsilon
y_{i}+u_{i}t\right)  \Theta_{\varepsilon}\left(  t\right)  ,\;i=1,\ldots,d.
\label{recoilmetric}%
\end{equation}
Here the suffix $0$ denotes temporal (Liouville) components, $\frac{u_{i}%
}{\varepsilon}\propto g_{s}\frac{\Delta k_{i}}{M_{s}}$, where $\Delta
k_{i}=\left(  k_{1}-k_{2}\right)  _{i}$ and $k_{1}\left(  k_{2}\right)  $ is
the momentum of the propagating closed-string state before (after) the recoil,
$y_{i}$ are the spatial collective coordinates of the D- particle and
$\varepsilon^{-2}$ is identified with the target Minkowski time $t$ for
$t\gg0$ after the collision; $\Theta_{\varepsilon}\left(  t\right)  $ is the
regularised step function represented by
\begin{equation}
\Theta_{\varepsilon}\left(  t\right)  =\frac{1}{2\pi i}\int_{-\infty}^{\infty
}\frac{dq}{q-i\varepsilon}e^{iqt},
\end{equation}
and $u_{i}$ is small,. For our purposes the Liouville and Minkowski times can
be identified. Now for large $t,$ to leading order,%
\begin{equation}
g_{0i}\simeq\overline{u}_{i}\equiv\frac{u_{i}}{\varepsilon} \label{finsler}%
\end{equation}
where $\Delta p_{i}$ is the momentum transfer during a collision and $M_{s}$
is the string mass scale, $g_{s}<1$ is the string coupling, assumed weak, and
the combination $M_{s}/g_{s}$ is the D-particle mass, playing the r\^{o}le of
the Quantum Gravity scale in this problem, i.e. the Planck mass; this
formalism has been used to establish a phenomenological model where the
couplings $u_{i}$ are taken to be stochastic and modeled by a gaussian
process\cite{Sarkar}. The stochasticity is due to two contributions. The
D-particle has an initial random recoil velocity due to capture of stringy
matter and subsequent release. Furthermore sumperimposed on this randomness
are quantum fluctuations due to vacuum string excitations.The process of
capture and emission does not have to conserve flavour. Consequently we need
to generalise the stochastic structure to allow for this.The fluctuations of
each component of the metric tensor $g^{\alpha\beta}$ will then not be just
given by the simple recoil distortion (\ref{recoilmetric}), but instead will be
taken to have a $2\times2$ (``flavour'') structure \cite{bernabeu}:
\begin{eqnarray}
  {g^{00}} &=& \left( { - 1 + {r_4}} \right)\underline{\underline 1}  \nonumber\\
  {g^{01}}&=& {g^{10}} = {r_0}\underline{\underline 1}  + {r_1}{\sigma _1} + {r_2}{\sigma _2} + {r_3}{\sigma _3} \nonumber\\
  {g^{11}}&=& \left( {1 + {r_5}} \right)\underline{\underline 1} \label{stochastic metric}
\end{eqnarray}
where $\underline{\underline 1}$ , is the identity operator and $\sigma_{i}$ are\ the Pauli matrices
and the $r_{\mu}$ are random variables which obey%
\begin{equation}
\left\langle r_{\mu}\right\rangle =0,\;\left\langle r_{\mu}r_{\nu
}\right\rangle =\Delta_{\mu}\delta_{\mu\nu}. \label{averages}%
\end{equation}
The above parametrisation has been taken for simplicity and also we will
consider motion to be in the $x$- direction which is natural since the meson
pairs move collinearly. The above parametrisation has been taken for
simplicity and we will also consider motion to be in the $x$- direction which
is natural since the meson pairs move collinearly. In any given realisation of
the random variables for a gravitational background,\ the system evolution can
be considered to be given by the Klein-Gordon equation for a two component
spinless neutral meson field $\Phi=\left(
\begin{array}
[c]{c}%
\phi_{1}\\
\phi_{2}%
\end{array}
\right)  $ (corresponding to the two flavours) with mass matrix $m=\frac{1}%
{2}\left(  m_{1}+m_{2}\right)  \underline{\underline 1}+\frac{1}{2}\left(  m_{1}%
-m_{2}\right)  \sigma_{3}$. We thus have
\begin{equation}
(g^{\alpha\beta}D_{\alpha}D_{\beta}-m^{2})\Phi=0 \label{KleinGordon}
\end{equation}
where $D_{\alpha}$ is the covariant derivative. Since the Christoffel symbols
vanish (as the $a_{i}$ are independent of space-time) the $D_{\alpha}$
coincide with $\partial_{\alpha}$. Hence within this flavour changing
background (\ref{KleinGordon}) becomes
\begin{equation}
\left(  g^{00}\partial_{0}^{2}+2g^{01}\partial_{0}\partial_{1}+g^{11}%
\partial_{1}^{2}\right)  \Phi-m^{2}\Phi=0. \label{KG2}
\end{equation}
We shall consider the two particle tensor product states made from the single
particle states $\left|  \uparrow^{\left(  i\right)  }\right\rangle $ and
$\left|  \downarrow^{\left(  i\right)  }\right\rangle $ $i=1,2$. This includes
the correlated states which include the state of the $\omega$ effect. In view
of (\ref{KG2}) the evolution of this state is governed by a hamiltonian
$\widehat{H}$ \
\begin{equation}
\widehat{H}=g^{01}\left(  g^{00}\right)  ^{-1}\widehat{k}-\left(
g^{00}\right)  ^{-1}\sqrt{\left(  g^{01}\right)  ^{2}{k}^{2}-g^{00}\left(
g^{11}k^{2}+m^{2}\right)  } \label{GenKG}%
\end{equation}
which is the natural generalisation of the standard Klein Gordon hamiltonian
in a one particle situation where $\widehat{k}\left|  \pm k,\alpha
\right\rangle =\pm k\left|  \pm k,\alpha\right\rangle $ with $\alpha=\uparrow$
or $\downarrow$. $\widehat{H}$ is a single particle hamiltonian and in order
to study two particle states associated with $i=1,2$ we can define
$\widehat{H}_{i}$ in the natural way (in terms of $\widehat{H}$) and then the
total hamiltonian is $\mathcal{H}=\sum_{i=1}^{2}$ $\widehat{H}_{i}$. The
effect of space-time foam on the initial entangled state of two neutral mesons
is conceptually difficult to isolate, given that the meson state is itself
entangled with the bath. Nevertheless, in the context of our specific model,
which is written as a stochastic hamiltonian, one can estimate the order of
the associated $\omega$-effect of \cite{bernabeu} by applying non-degenerate
perturbation theory to the states $\left|  \uparrow^{\left(  i\right)
}\right\rangle $, $\left|  \downarrow^{\left(  i\right)  }\right\rangle $,
$i=1,2$ (where the label $\pm k$ is redundant since $i$ already determines
it). Although it would be more rigorous to consider the corresponding density
matrices, traced over the unobserved gravitational degrees of freedom, in
order to obtain estimates it will suffice formally to work with single-meson
state vectors .

Owing to the form of the hamiltonian (\ref{GenKG}) the gravitationally
perturbed states will still be momentum eigenstates. The dominant features of
a possible $\omega$-effect can be seen from a term $\widehat{H_{I}}$ in the
interaction hamiltonian
\begin{equation}
\widehat{H_{I}}=-\left(  {r_{1}\sigma_{1}+r_{2}\sigma_{2}}\right)  \widehat{k}
\label{inthamil}%
\end{equation}
which is the leading order contribution in the small parameters $r_{i}$ (c.f.
(\ref{stochastic metric}),(\ref{GenKG})) in $H$ (i.e $\sqrt{\Delta_{i}}$ are
small). In first order non-degenerate perturbation theory the gravitational
dressing of $\left\vert {\downarrow}^{\left(  i\right)  }\right\rangle $ leads
to a state:%

\begin{equation}
\left\vert \downarrow^{\left(  i\right)  }\right\rangle _{QG}=\left\vert
\downarrow^{\left(  i\right)  }\right\rangle +\left\vert \uparrow^{\left(
i\right)  }\right\rangle \alpha^{\left(  i\right)  }%
\end{equation}
where
\begin{equation}
\alpha^{\left(  i\right)  }=\frac{\left\langle \uparrow^{\left(  i\right)
}\right\vert \widehat{H_{I}}\left\vert \downarrow^{\left(  i\right)
}\right\rangle }{E_{2}-E_{1}} \label{qgpert}%
\end{equation}
and correspondingly for $\left\vert {\uparrow}^{\left(  i\right)
}\right\rangle $ the dressed state is obtained from (\ref{qgpert}) by
$\left\vert \downarrow\right\rangle \leftrightarrow\left\vert \uparrow
\right\rangle $ and $\alpha\rightarrow\beta$ where
\begin{equation}
\beta^{\left(  i\right)  }=\frac{\left\langle \downarrow^{\left(  i\right)
}\right\vert \widehat{H_{I}}\left\vert \uparrow^{\left(  i\right)
}\right\rangle }{E_{1}-E_{2}} \label{qgpert2}%
\end{equation}
Here the quantities $E_{i}=(m_{i}^{2}+k^{2})^{1/2}$ denote the energy
eigenvalues, and $i=1$ is associated with the up state and $i=2$ with the down
state. With this in mind the totally antisymmetric \textquotedblleft
gravitationally-dressed\textquotedblright\ state can be expressed in terms of
the unperturbed single-particle states as:%

\begin{equation}%
\begin{array}
[c]{l}%
\left\vert {\uparrow}^{\left(  1\right)  }\right\rangle _{QG}\left\vert
{\downarrow}^{\left(  2\right)  }\right\rangle _{QG}-\left\vert {\downarrow
}^{\left(  1\right)  }\right\rangle _{QG}\left\vert {\uparrow}^{\left(
2\right)  }\right\rangle _{QG}=\\
\left\vert {\uparrow}^{\left(  1\right)  }\right\rangle \left\vert
{\downarrow}\right\rangle ^{\left(  2\right)  }-\left\vert {\downarrow
}\right\rangle ^{\left(  1\right)  }\left\vert {\uparrow}\right\rangle
^{\left(  2\right)  }\\
+\left\vert {\downarrow}^{\left(  1\right)  }\right\rangle \left\vert
{\downarrow}^{\left(  2\right)  }\right\rangle \left(  {\beta^{\left(
1\right)  }-\beta^{\left(  2\right)  }}\right)  +\left\vert {\uparrow
}^{\left(  1\right)  }\right\rangle \left\vert {\uparrow}^{\left(  2\right)
}\right\rangle \left(  {\alpha^{\left(  2\right)  }-\alpha^{\left(  1\right)
}}\right) \\
+\beta^{\left(  1\right)  }\alpha^{\left(  2\right)  }\left\vert {\downarrow
}^{\left(  1\right)  }\right\rangle \left\vert {\uparrow}^{\left(  2\right)
}\right\rangle -\alpha^{\left(  1\right)  }\beta^{\left(  2\right)
}\left\vert {\uparrow}^{\left(  1\right)  }\right\rangle \left\vert
{\downarrow}^{\left(  2\right)  }\right\rangle \\
\label{entangl}%
\end{array}
\end{equation}
It should be noted that for $r_{i}\propto\delta_{i1}$ the generated $\omega
$-like effect corresponds to the case  $\alpha^{\left(  i\right)  }=-\beta^{\left(  i\right)  }$, while the
$\omega$-effect of \cite{bernabeu1} (\ref{CPTV}) corresponds to $r_{i}%
\propto\delta_{i2}$ (and
$\alpha^{\left(  i\right)  }=\beta^{\left(  i\right)
}$). The velocity recoil
$u$ is given by
\[
u\sim r
\]
and so the variance $\Delta$ can be determined once the
variance of $u$ is known. In the perturbative approach we do not need to
specify the higher order moments provided they are finite and small. Rigour
would demand a density matrix approach but in order to estimate
the order of the effect it suffices to note that%
\begin{equation}
O\left(  \left\vert \omega^{2}\right\vert \right)  \sim\left\langle \left\vert
{\beta^{\left(  1\right)  }-\beta^{\left(  2\right)  }}\right\vert
^{2}\right\rangle \label{Estimate}%
\end{equation}
which is an estimate for $|\omega|^{2}$. It is necessary to examine the
moments for $u$.

\bigskip

\section{A string theory estimate for the variance of the
D-particle\label{sec:3}}

\bigskip

In order to estimate the variance of the recoil velcoity we need to understand
the dynamics of D-particle foam. It is imperative to consider D-particle
quantisation and in particular D-particle recoil as a result of scattering off
stringy matter. The issue of recoil is not fully understood and it is not our
purpose here to delve into the subtleties of recoil and operators describing
recoil. We will rather proceed on the basis of a proposal which has had some
success in the past \cite{kogan}. The energy of a D-particle is independent of
its position. Consequently in bosonic string theory there are 25 zero modes,
the Dirichlet directions. Zero modes lead to infrared divergences in loops in
a field theory setting where collective co-ordinates are used to isolate these
infrared divergences. This is essentially due to the naivety of the
perturbation series that is used and can be addressed using a coherent state
formalism. The situation is similar but, in some ways, worse for string theory
since the second quantised formalism for strings is more rudimentary. In
string theory the use of first quantisation requires a sum over Riemann sheets
with different moduli parameters. The formal transition from one surface to
another of lower genus (within string perturbation theory ) has singularities
associated with infrared divergences in the integration over moduli
parameters. In the case of a disc $D$ an incipient annulus $\Sigma$ can be
found by making two punctures and attaching a long thin strip (somewhat like
the strap in a handbag and dubbed wormholes). In the limit of vanishing width
this wormhole, is represented by a region in moduli space, integration over
which leads to an infrared divergence. A similar argument would apply to a
Riemann sphere where two punctures would be connected by a long thin tube. For
the disc the divergence will be due to propagation of zero mode open string
state while for the Riemann sphere it would be an analogous closed string
state along the wormhole. Let us examine this divergence explicitly. Consider
the correlation function $\left\langle V_{1}V_{2}V_{3}\ldots V_{n}%
\right\rangle _{\Sigma}$ for vertex operators \cite{strings} $\left\{
V_{i}\right\}  _{i=1,\ldots,n}$ where $\Sigma$ is the Riemann surface of an
annulus; it can be deformed into a disc with a wormhole attached
\cite{coll},\cite{szabo}. The set of vertex operators include necessarily any
higher dimensional D-branes necessary to conserve string charge. However such
vertex insertions clearly do not affect the infrared divergence caused by
\ wormwholes. The correlation function can be expressed as
\begin{equation}
\left\langle V_{1}V_{2}V_{3}\ldots V_{n}\right\rangle _{\Sigma}=\sum_{a}%
\int{\mathrm{{d}}s_{1}}\int{\mathrm{{d}}s_{2}}\int\frac{\mathrm{d}q}%
{q}q^{h_{a}-1}\left\langle \phi_{a}\left(  s_{1}\right)  \phi_{a}\left(
s_{2}\right)  V_{1}V_{2}V_{3}\ldots V_{n}\right\rangle _{D} \label{infrared}%
\end{equation}
where $s_{1}$ and $s_{2}$ are the positions of the punctures on $\partial
D$.The $\left\{  \phi_{a}\right\}  $ are a complete set of eigenstates of the
Virasoro operator $L_{0}$ with conformal weights $h_{a}$\cite{mathieu} and $q$
is a Teichmuller parameter associated with the added thin strip. Clearly there
is a potential divergence associated with its disappearance $q\rightarrow0$
and this corresponds to a long thin strip attached to the disc. For a static
D-particle the string co-ordinates $\overrightarrow{X}_{D}=\left\{
X^{1},X^{2},X^{3},\ldots,X^{25}\right\}  $ have Dirichlet boundary conditions
while $X^{0}$ has Neumann boundary conditions (in the static gauge)%
\begin{equation}
\left.  \frac{\partial}{\partial\sigma}X^{0}\left(  \tau,\sigma\right)
\right|  _{\sigma=0}=0=\left.  \frac{\partial}{\partial\sigma}X^{0}\left(
\tau,\sigma\right)  \right|  _{\sigma=\pi}%
\end{equation}
where $\left(  \tau,\sigma\right)  $ is a co-ordinisation of the
worldsheet.The associated translational zero mode is given by%
\begin{equation}
\phi^{i}\left(  X,\omega\right)  =\frac{\sqrt{g_{s}}}{4}\partial_{n}X_{D}%
^{i}\e^{i\omega X^{0}}, \label{zeromode}%
\end{equation}
where $\partial_{n}$ denotes a derivative in the $X^{i}$ Dirichlet direction,
and is an element of the set $\left\{  \phi_{a}\right\}  $. The conformal
weight $h_{i}=1+\alpha^{\prime}\omega^{2}$. The relevant part of the integral
in (\ref{infrared}) is
\begin{equation}
\int_{0}^{1}dq\int_{-\infty}^{\infty}d\omega\,q^{-1+\alpha^{\prime}\omega^{2}%
}=\int_{0}^{1}dq\,\frac{1}{q\left(  -\log q\right)  ^{1/2}} \label{relevant}%
\end{equation}
and is divergent because of the behaviour of the integrand near $q=0.$ This
can be regularized by putting a lower cut-off $q>\delta\rightarrow0$ in the
integral.The correlation function on $\Sigma$ can be computed to be
\cite{coll}
\begin{equation}
\left\langle V_{1}V_{2}V_{3}\ldots V_{n}\right\rangle _{\Sigma}=-\frac{g_{s}%
}{16T}\log\delta\left\langle \partial_{n}\overrightarrow{X}_{D}\left(
s_{1}\right)  .\partial_{n}\overrightarrow{X}_{D}\left(  s_{2}\right)
V_{1}V_{2}V_{3}\ldots V_{n}\right\rangle _{D} \label{infred}%
\end{equation}
where $T$ is a cut-off for large (target) time. Division by it removes
divergencies due to the integration over the world-sheet zero modes of the
target time. These should not be confused with divergencies associated with
pinched world-sheet surfaces, proportional to $\mathrm{log}\delta$, that we
are interested in here. These latter divergences cause conventional conformal
invariance to fail.

However, in one approach, it was argued sometime ago, that these divergences
can be canceled if D-particles are allowed to recoil~\cite{kogan,szabo}, as a
result of momentum conservation during their scattering with string states. In
fact, in our D-particle foam model~\cite{Dfoam}, this is not a simple
scattering process, as it involves capture and re-emission of the string state
by the D-particle defect. In simple terms, this process involves splitting of
strings by the defect. In a world-sheet (first quantization) framework, such
processes are described by appropriate vertex operators, whose operator
product expansion close on a (local) logarithmic algebra~\cite{kogan,szabo}.
The translational zero modes are associated with infinitesimal translations in
the Dirichlet directions. Given that D-particles do not have any internal or
rotational degrees of freedom, these modes should give us valuable information
concerning recoil. Moreover (for $\omega=0$) there is a degeneracy in the
conformal weights between $\phi^{i}$, $\partial_{\omega}\phi^{i}$ and the
identity operator and also the conformal blocks in the corresponding algebra
have logarithmic terms.

To understand the formal structure of the world-sheet deformation operators
pertinent to the recoil/capture process, we first notice that the world-sheet
boundary operator ${V_D}$ describing the excitations of a
moving heavy D0-brane is given in the tree approximation by:

\[{V_D} = \int\limits_{\partial D} {\left( {{y_i}{\partial _n}{X^i} + {u_i}{X^0}{\partial _n}{X^i}} \right) \equiv \int\limits_{\partial D} {{Y_i}\left( {{X^0}} \right){\partial _n}{X^i}} } \label{recoilop}\]
where $u_{i}$ and $y_{i}$ are the velocity \ and position of the D-particle
respectively and $Y_{i}\left(  X^{0}\right)  \equiv y_{i}+u_{i}X^{0}$. To
describe the capture/recoil we need an operator which has non-zero matrix
elements between different states of \ the D-particle and is turned on
``abruptly'' in target time. One way of doing this is to put~\cite{kogan} a
$\Theta\left(  X^{0}\right)  $, the Heavyside function, in front of
${V_D}$ which models an impulse whereby the D-particle starts
moving at $X^{0}=0$. Using Gauss's theorem this impulsive ${V_D}$, denoted by ${V_D}^{imp}$, can be represented
as
\begin{equation}
{V_D}^{imp}=\sum_{i=1}^{25}\int_{D}d^{2}z\,\partial_{\alpha
}\left(  \left[  u_{i}X^{0}\right]  \Theta\left(  X^{0}\right)  \partial
^{\alpha}X^{i}\right)  =\sum_{i=1}^{25}\int_{\partial D}d\tau\,u_{i}%
X^{0}\Theta\left(  X^{0}\right)  \partial_{n}X^{i}. \label{fullrec}%
\end{equation}
Since $X^{0}$ is an operator it will be necessary to define $\Theta\left(
X^{0}\right)  $ as an operator using the contour integral%
\begin{equation}
\Theta_{\varepsilon}\left(  X^{0}\right)  =-\frac{i}{2\pi}\int_{-\infty
}^{\infty}\frac{d\omega}{\omega-i\varepsilon}\e^{i\omega X^{0}}
\end{equation}
 with $\varepsilon\rightarrow0+$.

Hence we can consider%
\begin{equation}
\label{Depsilonop}D_{\varepsilon}(X^{0}) \equiv D (X^{0} ; \varepsilon) =
X^{0}\Theta_{\varepsilon}\left(  X^{0}\right)  =-\int_{-\infty}^{\infty}%
\frac{d\omega}{\left(  \omega-i\varepsilon\right)  ^{2}}\e^{i\omega X^{0}}~.
\end{equation}
The introduction of the feature of impulse in the operator breaks conventional
conformal symmetry, but a modified logarithmic conformal algebra holds. A
generic logarithmic algebra in terms of operators $\mathcal{C}$ and
$\mathcal{D}$ and the stress tensor $T\left(  z\right)  $\ (in complex tensor
notation ) satisfies the operator product expansion%

\[\begin{array}{l}
 T\left( z \right)C\left( {w,\overline w } \right) \sim \frac{\Delta }{{{{\left( {z - w} \right)}^2}}}C\left( {w,\overline w } \right) + \frac{{\partial C\left( {w,\overline w } \right)}}{{{{\left( {z - w} \right)}^2}}} +  \cdots  \\
 T\left( z \right)D\left( {w,\overline w } \right) \sim \frac{\Delta }{{{{\left( {z - w} \right)}^2}}}D\left( {w,\overline w } \right) + \frac{1}{{{{\left( {z - w} \right)}^2}}}C\left( {w,\overline w } \right) + \frac{{\partial D\left( {w,\overline w } \right)}}{{{{\left( {z - w} \right)}^2}}} +  \cdots  \\
 \end{array}\]

and

\begin{eqnarray}
 \left\langle {C\left( {z,\overline z } \right)C\left( {0,0} \right)} \right\rangle  & \sim & 0, \label{can1} \\
 \left\langle {C\left( {z,\overline z } \right)D\left( {0,0} \right)} \right\rangle  & \sim & \frac{c}{{{{\left| z \right|}^{2\Delta }}}}, \label{can2} \\
 \left\langle {D\left( {z,\overline z } \right)D\left( {0,0} \right)} \right\rangle  & \sim & \frac{c}{{{{\left| z \right|}^{2\Delta }}}}\left( {\log \left| z \right| + d} \right), \label{can3}
 \end{eqnarray}

where $d$ is a constant. Since the conformal dimension of
$\e^{iqX^{0}}$ is $\frac{q^{2}}{2}$ we find that

\begin{equation}
T\left(  w\right)  D_{\varepsilon}\left(  z\right)  \sim-\frac{\varepsilon
^{2}}{2\left(  w-z\right)  ^{2}}D_{\varepsilon}\left(  z\right)  +\frac
{1}{\left(  w-z\right)  ^{2}}\varepsilon\Theta_{\varepsilon}\left(
X^{0}\right)  +\cdots
\end{equation}

and so a logarithmic conformal algebra structure arises if we define

\begin{equation}
\label{Cepsilonop}C_{\varepsilon} (X^{0}) \equiv C(X^{0} ; \varepsilon)
=\varepsilon\Theta_{\varepsilon}\left(  X^{0}\right)  ~,
\end{equation}

suppressing, for simplicity, the non-holomorphic piece. The above logarithmic
conformal field theory structure is found with this identification. Similarly
we find%
\[
T\left(  w\right)  C_{\varepsilon}\left(  z\right)  \sim-\frac{\varepsilon
^{2}}{2\left(  w-z\right)  ^{2}}C_{\varepsilon}\left(  z\right)  +\cdots
\]
Consequently $\Delta$ for $C_{\varepsilon}\left(  z\right)  $ and
$D_{\varepsilon}\left(  z\right)  $ is $-\frac{\varepsilon^{2}}{2}$. A
calculation (in a euclidean metric) for a disc of size $L$ with a
short-distance worldsheet cut-off $a$ reveals that as $\varepsilon
\rightarrow0$%

\begin{eqnarray}
 \left\langle {{C_\varepsilon }\left( {z,\overline z } \right){C_\varepsilon }\left( {0,0} \right)} \right\rangle  & \sim & O\left( {{\varepsilon ^2}} \right), \label{canep1} \\
 \left\langle {{C_\varepsilon }\left( {z,\overline z } \right){D_\varepsilon }\left( {0,0} \right)} \right\rangle  & \sim & \frac{\pi }{2}\sqrt {\frac{\pi }{{{\varepsilon ^2}\alpha }}} \left( {1 - 4{\varepsilon ^2}\log \left| {\frac{z}{a}} \right|} \right), \label{canep2} \\
 \left\langle {{D_\varepsilon }\left( {z,\overline z } \right){D_\varepsilon }\left( {0,0} \right)} \right\rangle  & \sim & \frac{\pi }{2}\sqrt {\frac{\pi }{{{\varepsilon ^2}\alpha }}} \left( {\frac{1}{{{\varepsilon ^2}}} - 4\log \left| {\frac{z}{a}} \right|} \right), \label{canep3}
\end{eqnarray}

where $\alpha=\log\left\vert \frac{L}{a}\right\vert ^{2}$. We consider
$\varepsilon\rightarrow0+$ such that
\begin{equation}
\varepsilon^{2}\alpha\sim\frac{1}{2\eta}=O\left(  1\right)  ~,
\label{epscutoff}%
\end{equation}
where $\eta$ is the time signature and the right-hand side is kept fixed as
the cutoff runs; it is then straightforward to see that (\ref{canep1}),
(\ref{canep2}), and (\ref{canep3}) are consistent with (\ref{can1}),
(\ref{can2}), and (\ref{can3}). It is only under the condition
(\ref{epscutoff}) that the recoil operators $C_{\varepsilon}$ and
$D_{\varepsilon}$ obey a closed logarithmic conformal algebra~\cite{kogan}:

\begin{eqnarray}
 \left\langle {{C_\varepsilon }\left( z \right){C_\varepsilon }\left( 0 \right)} \right\rangle  & \sim & 0 \\
 \left\langle {{C_\varepsilon }\left( z \right){D_\varepsilon }\left( 0 \right)} \right\rangle  & \sim & 1 \\
 \left\langle {{D_\varepsilon }\left( z \right){D_\varepsilon }\left( 0 \right)} \right\rangle  & \sim &  - 4\eta \log \left| {\frac{z}{L}} \right| \label{CD}
 \end{eqnarray}

The reader should notice that the full recoil operators, involving
$\partial_{n}X^{i}$ holomorphic pieces with the conformal-dimension-one
entering (\ref{fullrec})), obey the full logarithmic algebra (\ref{can1}),
(\ref{can2}), (\ref{can3}) with conformal dimensions $\Delta=1-\frac
{\varepsilon^{2}}{2}$. From now on we shall adopt the euclidean signature
$\eta=1$.

We next remark that, at tree level in the string perturbation sense, the
stringy sigma model (inclusive of the D-particle \ boundary term and other
\ vertex operators) is a two dimensional renormalizable quantum field theory;
hence for generic couplings $g^{i}$ it is possible to see how the couplings
run in the renormalization group sense with changes in the short distance
cut-off through the beta functions $\beta^{i}$. In the world-sheet
renormalization group \cite{klebanov}, based on expansions in powers of the
couplings, $\beta^{i}$ has the form ( with no summation over the repeated
indices)
\begin{equation}
\beta^{i}=y_{i}g^{i}+\ldots\label{beta fn}%
\end{equation}
where $y_{i}$ is the anomalous dimension, which is related to the conformal
dimension $\Delta_{i}$ by $y_{i} = \Delta_{i} - \delta$, with $\delta$ the
engineering dimension (for the holomorphic parts of vertex operators for the
open string $\delta= 1$). The $\ldots$ in (\ref{beta fn}) denote higher orders
in $g^{i}$. Consequently, in our case, we note that the (renormalised)
D-particle recoil velocities $u^{i}$ constitute such $\sigma$-model couplings,
and to lowest order in the renormalised coupling $u_{i}$ the corresponding
$\beta$ function satisfies%
\begin{equation}
\frac{du^{i}}{d\log\Lambda}=-\frac{\varepsilon^{2}}{2}u^{i}. \label{rengp}%
\end{equation}
where $\Lambda$ is a (covariant) world-sheet renormalization-group scale. In
our notation, we identify the logarithm of this scale with $\alpha
=\log\left\vert \frac{L}{a}\right\vert ^{2}$, satisfying (\ref{epscutoff}).

An important comment is now in order concerning the interpretation of the flow
of this world-sheet renormalization group scale as a target-time flow. The
target time $t$ is identified through $t=2\log\Lambda$. For completeness we
recapitulate the arguments of \cite{kogan} leading to such a conclusion. Let
one make a scale transformation on the size of the world-sheet
\begin{equation}
L\rightarrow L^{\prime}=\e^{t/4}L \label{fsscaling}%
\end{equation}
which is a finite-size scaling (the only one which has physical sense for the
open string world-sheet). Because of the relation between $\varepsilon$ and
$L$ (\ref{epscutoff}) this transformation will induce a change in
$\varepsilon$
\begin{equation}
\varepsilon^{2}\rightarrow\varepsilon^{\prime2}=\frac{\varepsilon^{2}%
}{1+\varepsilon^{2}t} \label{epsilontransform}%
\end{equation}
(note that if $\varepsilon$ is infinitesimally small, so is $\varepsilon
^{\prime}$ for any finite $t$). From the scale dependence of the correlation
functions (\ref{CD}) that $C_{\varepsilon}$ and $D_{\varepsilon}$ transform
as ${C_\varepsilon } \to {C_{\varepsilon '}} = {C_\varepsilon }$ and
${D_\varepsilon } \to {D_{\varepsilon '}} = {D_\varepsilon } + t{C_\varepsilon }$.

Hence the coupling constants in front
of $C_{\varepsilon}$ and $D_{\varepsilon}$ in the recoil operator
(\ref{recoilop}), i.e. the velocities $u_{i}$ and spatial collective
coordinates $y_{i}$ of the brane, must transform like:
\begin{equation}
u_{i}\rightarrow u_{i}~~,~~y_{i}\rightarrow y_{i}+u_{i}t \label{scale2}%
\end{equation}
This is nothing other than the Galilean transformation for the
heavy D-particles and thus it demonstrates that the finite size scaling
parameter $t$, entering (\ref{fsscaling}), plays the r\^{o}le of target time,
on account of (\ref{epscutoff}). Notice that (\ref{scale2}) is derived upon
using (\ref{CD}), that is in the limit where $\varepsilon\rightarrow0$. This
will become important later on, where we shall discuss (stochastic) relaxation
phenomena in our recoiling D-particle.

Thus, in the presence of recoil a world-sheet scale transformation leads to an
evolution of the $D$-brane in target space, and from now on we identify the
world-sheet renormalization group scale with the target time $t$. In this
sense, equation (\ref{rengp}) is an evolution equation in target time.

However, this equation does not capture quantum-fluctuation aspects of $u^{i}$
about its classical trajectory with time $u_{i}(t)$. Going to higher orders in
perturbation theory of the quantum field theory at fixed genus does not
qualitatively alter the situation in the sense that the equation remains
deterministic. However the effect of string perturbation theory where higher
genus surfaces are considered and re-summed in some appropriate limits can be
shown to give rise to quantum fluctuations in $u^{i}$ \cite{SarkarNonExtStat}
which leads to a Langevin equation \cite{gardiner2} rather than (\ref{rengp}). The equation has
the form
\begin{equation}
\frac{d{\bar{u}}^{i}}{dt}=-\frac{1}{4t}{\bar{u}}^{i}\,+\frac{g_{s}}%
{\sqrt{2\alpha^{\prime}}}t^{1/2}\xi\left(  t\right)  \label{Langevin}%
\end{equation}
where $t=\varepsilon^{-2}$ and $\xi\left(  t\right)  $ represents white noise.
This equation is valid for large $t$. From the above analysis it is known that
\cite{szabo} (c.f. Appendix A)) that to $O\left(  g_{s}^{2}\right)  $ the
correlation for $\xi\left(  t\right)  $ is $\bar{u}^{i}$ independent, \ and
for time scales of interest, is correlated like white noise ; hence the
correlation of $\xi\left(  t\right)  $ has the form:
\begin{equation}
\left\langle \xi\left(  t\right)  \xi\left(  t^{\prime}\right)  \right\rangle
=\delta\left(  t-t^{\prime}\right)  . \label{noise}%
\end{equation}
Since the vectorial nature of $\bar{u}^{i}$ is not crucial for our analysis, we
will suppress it and consider the single variable $\bar{u}$. Moreover the mean of $\bar{u} $   vanishes.

\bigskip

\section{Langevin equations and fluctuating string coupling \label{sec:4}}
It is possible to solve \cite{SarkarNonExtStat} the Langevin equation (\ref{Langevin}) (with (\ref{noise})).The quantity of interest which is an important input for our estimate of the omega effect is the variance of $u$ the velocity of the D-particle. If the D-particle is typically interacting with matter on a time scale of $t_0$, then the effect of a large number of such collisions can be calculated by performing an ensemble average over a distribution of $u_0$ the initial velocity at the time of capture; a gaussian distribution with zero mean and variance $\sigma$ is assumed for $u_0$ and is appropriate in the absence of a priori knowledge.Using generic properties of strings \cite{emnuncertnew} consistent with the space-time uncertainties \cite{yoneya}, the capture and re-emission time $t_0$, involves the growth of a stretched string between the D-particle and the D-brane world  (e.g. a D3 brane corresponding to our 3-dimensional world). The minimisation of the energy associated with intermediate string tension and oscillations  leads to :
\[
t_0  \sim \alpha 'p^0
\label{capturetime}
\]
where $p^0$ is the energy of the intermediate string state. For neutral kaons it is appropriate to consider non-relativistic particles and so
$p^0  \sim \sqrt {m_1 ^2  + m_2 ^2 } $.  This result is not modified even if there are more than $2$ D-particles in close proximity. Using this we can infer \cite{SarkarNonExtStat} from the solution of the Fokker-Planck equation associated with Eqn.(\ref{Langevin}) that
\[{\mathop{\rm var}} \left( u \right) \simeq \frac{{2\left( {1 + \sqrt 2 } \right)\left[ {g_s^2\alpha 'p_0^2 + 15{\sigma ^2}} \right]}}{{15}}
\label{nonext}
\]
for the variance of $u$ due to capture of stringy matter by D particles. The terms of order $\sigma ^2$ have not been written explicitly for several reasons. In most models of quantum gravity, we expect that
${\sigma ^2} \le g_s^2\alpha 'p_0^2$ since it is a natural assumption to make for a dispersion due to (quantum) fluctuations of the recoil velocity of heavy D-particles of average mass  ${M_s}/{g_s}$.We now return to the estimate for the $\omega$ effect \ref{Estimate}. It is straightforward to show that
\[{\beta ^{\left( 1 \right)}} = k\frac{{\left( {r_1^{\left( 1 \right)} + ir_2^{\left( 1 \right)}} \right)}}{{{E_1} - {E_2}}}\] and
\[{\beta ^{\left( 2 \right)}} =  - k\frac{{\left( {r_1^{\left( 2 \right)} + ir_2^{\left( 2 \right)}} \right)}}{{{E_1} - {E_2}}}.\]
Hence
\[\left\langle {{{\left| {{\beta ^{\left( 1 \right)}} - {\beta ^{\left( 2 \right)}}} \right|}^2}} \right\rangle  = 2{k^2}\frac{{\left\langle {r_1^{\left( 1 \right)\,2} + r_2^{\left( 1 \right)\,2}} \right\rangle }}{{{{\left( {{E_1} - {E_2}} \right)}^2}}}
\label{Estimate2}
\]
on noting that
\[\begin{array}{l}
 \left\langle {r_1^{\left( 1 \right)\,}r_1^{\left( 2 \right)\,}} \right\rangle  = \left\langle {r_2^{\left( 1 \right)\,}r_2^{\left( 2 \right)\,}} \right\rangle  = 0 \\
 \left\langle {r_i^{\left( 1 \right)2\,}} \right\rangle  = \left\langle {r_i^{\left( 2 \right)2\,}} \right\rangle  = \Delta {}_i,\;i = 1,2. \\
 \end{array}\]
This embodies the independence of the random variables which differ in their indices. Furthermore at this order in perturbation theory no information is required about higher order moments (provided they remain bounded). From Eqn.(\ref{Estimate}) and Eqn.(\ref{nonext}) we can conclude that
\[O\left( {\left| {{\omega ^2}} \right|} \right) \sim \frac{{2g_s^2{k^2}}}{{M_s^2}}\frac{{m_1^2 + m_2^2}}{{{{\left( {{m_1} - {m_2}} \right)}^2}}} = \frac{{g_s^2{k^2}}}{{M_s^2}}\frac{{{\zeta^2}{k^2}}}{{{{\left( {{m_1} - {m_2}} \right)}^2}}}
\label{Estimate3}
\]
where ${\zeta ^2} = \frac{{2\left( {m_1^2 + m_2^2} \right)}}{{{k^2}}}$. In
the estimate given in \cite{bernabeu} $\zeta$ was undetermined. It is a
success of this analysis that it is possible to obtain a simple formula for
it. On remembering the restrictions $r_{i}\propto\delta_{i1}$ or $r_{i}%
\propto\delta_{i2}$ discussed earlier, we find for kaons that $\zeta=O\left(  1\right)  $ and $\left|  \omega\right|
\sim2\times10^{-5}$ . Strictly speaking, since the constituent quarks making
up the kaons are charged, it is the gluons of the strong force which interact
with the D particles. However we shall not pursue this issue here but assume,
simplistically, that the kaon is an elementary neutral excitation. The present
bound for $\left|  \omega\right|  $ from DA$\Phi$NE is $\left|  \omega\right|
<10^{-3}$ with 95\% confidence level. Hence the the strength of the effect is
not far below the sensitivity of current experiments and may be detectable
when DA$\Phi$NE is upgraded. A similar analysis may be done for B mesons
produced in Super B factories.

\section{The Thermal Master Equation\label{sec:5}}
Master equations with a thermal bath have been argued to be relevant to
decoherence with space time foam \cite{Garay}. His argument was based on
problems of measurement \cite{giulini}. Also, by considering the infinite
redhshifting near the horizon for an observer far away from the horizon of a
black hole, Padmanabhan \cite{padmanabhan} has argued that a foam consisting
of virtual black holes would magnify Planck scale physics for observers
asymptotically far from the horizon and thus an effective non-local field
theory description should emerge. Garay has argued that mildest non-locality,
bilocality, can be \ rewritten with the help of a local auxilary field and
this plausibly leads to a description of space-time foam as a heat bath. A
thermal field represents a bath about which there is minimal information since
only the mean energy of the bath is known, a situation valid also for
space-time foam. In applications of quantum information it has been shown that
a system of two qubits (or two-level systems) initially in a separable state
(and interacting with a thermal bath) can actually be entangled by such a
single mode bath \cite{entanglement}. As the system evolves the degree of
entanglement is sensitive to the initial state. The close analogy between
two-level systems and neutral meson systems, together with the modelling by a
phenomenological thermal bath of space time foam, makes the study of thermal
master equations a rather intriguing one for the generation of $\omega$. The
hamiltonian $\mathcal{H}$ representing the interaction of \ two such two level
atoms with a single mode thermal field is
\begin{equation}
\mathcal{H}=\hbar\nu a^{\dag}a+\frac{1}{2}\hbar\Omega\sigma_{3}^{\left(
1\right)  }+\frac{1}{2}\hbar\Omega\sigma_{3}^{\left(  2\right)  }+\hbar
\gamma\sum_{i=1}^{2}\left(  a\sigma_{+}^{\left(  i\right)  }+a^{\dag}%
\sigma_{-}^{\left(  i\right)  }\right)  \label{thermal}%
\end{equation}
where $a$ is the bosonic annihilation operator for the mode of the thermal
field and the $\sigma$'s are the standard Pauli matrices for the 2 level
systems.The operators $a$ and $a^{\dag}$ satisfy
\begin{equation}
\left[  a,a^{\dag}\right]  =1,\left[  a^{\dag},a^{\dag}\right]  =\left[
a,a\right]  =0.
\end{equation}
The superscripts label the particle. This hamiltonian is known as the
Jaynes-Cummings hamiltonian \cite{Jaynes} and explicitly incorporates the back
reaction or entanglement between system and bath. This is in contrast with the
Lindlblad model and the Liouville stochastic metric model. In common with the
Lindblad model it is non-geometric. In the former the entanglement with the
bath has been in principle integrated over while in the latter it is
represented by a stochastic effect. An important feature of \ $\mathcal{H}$ in
\ref{thermal} is the block structure of subspaces that are left invariant by
$\mathcal{H}$. It is straightforward to show that the family of invariant
irreducible spaces $\mathcal{E}_{n}$ may be defined by $\left\{  \left\vert
e_{i}^{\left(  n\right)  }\right\rangle ,i=1,\ldots,4\right\}  $ where ( in
obvious notation, with $n$ denoting the number of oscillator quanta)
\begin{equation}
\left\vert e_{1}^{\left(  n\right)  }\right\rangle   \equiv\left\vert
\uparrow^{\left(  1\right)  },\,\uparrow^{\left(  2\right)  },n\right\rangle
,\label{equation1}\\
\left\vert e_{2}^{\left(  n\right)  }\right\rangle  \equiv\left\vert
\uparrow^{\left(  1\right)  },\,\downarrow^{\left(  2\right)  }%
,n+1\right\rangle ,\nonumber\\
\left\vert e_{3}^{\left(  n\right)  }\right\rangle   \equiv\left\vert
\downarrow^{\left(  1\right)  },\,\uparrow^{\left(  2\right)  }%
,n+1\right\rangle ,\nonumber
\end{equation}
and
\[
\left\vert e_{4}^{\left(  n\right)  }\right\rangle \equiv\left\vert
\downarrow^{\left(  1\right)  },\,\downarrow^{\left(  2\right)  }%
,n+2\right\rangle .
\]
The total space of states is a direct sum of the $\mathcal{E}_{n}$ for
different $n$. We will write $\mathcal{H}$ as $\mathcal{H}_{0}+\mathcal{H}%
_{1}$ where
\begin{equation}
\mathcal{H}_{0}=\nu a^{\dag}a+\frac{\Omega}{2}\left(  \sigma_{3}^{\left(
1\right)  }+\sigma_{3}^{\left(  2\right)  }\right)  \label{hamiltonian0}%
\end{equation}
and
\begin{equation}
\mathcal{H}_{1}=\gamma^{\prime}\sum_{i=1}^{2}\left(  a\sigma_{+}^{\left(
i\right)  }+a^{\dag}\sigma_{-}^{\left(  i\right)  }\right)  .
\label{hamiltonianI}%
\end{equation}
$n$ is a quantum number and gives the effect of the random environment. In our
era the strength $\gamma$ of the coupling with the bath is weak. We expect
heavy gravitational degrees of freedom and so $\Omega\gg\nu$. It is possible
to associate both thermal and highly non-classical density matrices with the
bath state.\ Nonethelss because of the block structure it can be shown
completely non-perturbatively that the $\omega$ contribution to the density
matrix is absent \cite{sarkar2}.

\bigskip We will calculate the stationary states in $\mathcal{E}_{n}$, using
degenerate perturbation theory, where appropriate. We will be primarily
interested in the dressing of the degenerate states $\left|  e_{2}^{\left(
n\right)  }\right\rangle $ and $\left|  e_{3}^{\left(  n\right)
}\right\rangle $ because it is these which contain the neutral meson entangled
state . In 2nd order perturbation theory the dressed states are
\begin{equation}
\left|  \psi_{2}^{\left(  n\right)  }\right\rangle =\left|  e_{2}^{\left(
n\right)  }\right\rangle -\left|  e_{3}^{\left(  n\right)  }\right\rangle
+O\left(  \gamma^{\prime\,3}\right)  \label{dressed1}%
\end{equation}
with energy $E_{2}^{\left(  n\right)  }=\left(  n+1\right)  \nu+O\left(
\gamma^{\prime\,3}\right)  $ and
\begin{equation}
\left|  \psi_{3}^{\left(  n\right)  }\right\rangle =\left|  e_{2}^{\left(
n\right)  }\right\rangle +\left|  e_{3}^{\left(  n\right)  }\right\rangle
+O\left(  \gamma^{\prime\,3}\right)  \label{dressed2}%
\end{equation}
with energy $E_{3}^{\left(  n\right)  }=\left(  n+1\right)  \nu+\frac
{2\gamma^{2}}{\Omega-\nu}+O\left(  \gamma^{\prime\,3}\right)  $. It is
$\left|  \psi_{2}^{\left(  n\right)  }\right\rangle $ which can in principle
give the $\omega$-effect. More precisely we would construct the state
$Tr\left(  \left|  \psi_{2}^{\left(  n\right)  }\right\rangle \left\langle
\psi_{2}^{\left(  n\right)  }\right|  \rho_{B}\right)  $ where $\rho_{B}$ is
the bath density matrix (and has the form $\rho_{B}=\sum_{n,m}p_{nn^{\prime}%
}\left|  n\right\rangle \left\langle n^{\prime}\right|  $ for suitable choices
of $p_{nn^{\prime}}$). To this order of approximation $\left|  \psi
_{2}^{\left(  n\right)  }\right\rangle $ cannot generate the $\omega$-effect
since there is no admixture of $\left|  e_{1}^{\left(  n\right)
}\right\rangle $ and $\left|  e_{4}^{\left(  n\right)  }\right\rangle $.
However it is a priori possible that this may change when higher orders in
$\gamma$. We can show that $\left|  \psi_{2}^{\left(  n\right)  }\right\rangle
=\left|  e_{2}^{\left(  n\right)  }\right\rangle -\left|  e_{3}^{\left(
n\right)  }\right\rangle $ to \textit{all orders} in $\gamma$ by directly
considering the hamiltonian matrix $\mathcal{H}^{\left(  n\right)  }$ for
$\mathcal{H}$ within $\mathcal{E}_{n}$; it is given by
\begin{equation}
\mathcal{H}^{\left(  n\right)  }=\left(
\begin{array}
[c]{cccc}%
\Omega+n\nu & \gamma\sqrt{n+1} & \gamma\sqrt{n+1} & 0\\
\gamma\sqrt{n+1} & \left(  n+1\right)  \nu & 0 & \gamma\sqrt{n+2}\\
\gamma\sqrt{n+1} & 0 & \left(  n+1\right)  \nu & \gamma\sqrt{n+2}\\
0 & \gamma\sqrt{n+2} & \gamma\sqrt{n+2} & \left(  n+2\right)  \nu-\Omega
\end{array}
\right)  . \label{matrix}%
\end{equation}
We immediately notice that
\begin{equation}
\mathcal{H}^{\left(  n\right)  }\left(
\begin{array}
[c]{c}%
0\\
1\\
-1\\
0
\end{array}
\right)  =\left(  n+1\right)  \nu\left(
\begin{array}
[c]{c}%
0\\
1\\
-1\\
0
\end{array}
\right)  \label{eigenvectoreqn}%
\end{equation}
and so, to all orders in $\gamma$, the environment does \textit{not} dress the
state of interest to give the $\omega$-effect; clearly this is independent of
any choice of $\rho_{B}$.

As we noted this block structure of the Hilbert space is a consequence of the
structure of $\mathcal{H}_{1}$ which is commonly called the hamiltonian in the
`rotating wave' approximation\cite{PikeSark}. (This nomenclature arises within
the context of quantum optics where this model is used extensively.) It is
natural to expect that, once the rotating wave approximation is abandoned, our
conclusion would be modified. We shall examine whether this expectation
materialises by adding to $\mathcal{H}$ a `non-rotating wave' piece
$\mathcal{H}_{2}$
\begin{equation}
\mathcal{H}_{2}=\gamma\sum_{i=1}^{2}\left(  a\sigma_{-}^{\left(  i\right)
}+a^{\dag}\sigma_{+}^{\left(  i\right)  }\right)  . \label{nonrotwave}%
\end{equation}
$\mathcal{H}_{2}$ does not map $\mathcal{E}_{n}$ into $\mathcal{E}_{n}$ as can
be seen from
\[
\mathcal{H}_{2}\left\vert e_{2}^{\left(  n\right)  }\right\rangle
=\gamma^{\prime}\left(  \sqrt{n+2}\,\left\vert e_{1}^{\left(  n+2\right)
}\right\rangle +\sqrt{n+1}\left\vert e_{4}^{\left(  n-2\right)  }\right\rangle
\right)  .
\]
We have noted that $\left\vert e_{3}^{\left(  n\right)  }\right\rangle
-\left\vert e_{2}^{\left(  n\right)  }\right\rangle $ is an eigenstate of
\ $\mathcal{H}_{0}+\mathcal{H}_{1}$. However the perturbation $\mathcal{H}%
_{2}$ \textit{annihilates} $\left\vert e_{3}^{\left(  n\right)  }\right\rangle
-\left\vert e_{2}^{\left(  n\right)  }\right\rangle $ and so this eigenstate
does not get dressed by $\mathcal{H}_{2}$. Hence the $\omega$ effect cannot be
rescued by moving away from the rotating wave scenario.

We should comment that a variant of the bath model based on non-bosonic
operators $a$ can be considered. \ This departs somewhat from the original
philosophy of Garay but may have some legitimacy from analysis in M-theory.
From an infinite momentum frame study it was suggested that D-particles may
satisfy infinite statistics\cite{InfStat}. Of course these D-particles would
be light contrary to the heavy D-particles that we have been considering and
so we will consider a more general deformation of the usual statistics.. If
flavour changes can arise due to D-particle interactions then it may be
pertinent to reinterpret the hamiltonian in this way. Generalised statistics
is characterised conventionally by a c-number deformation\cite{macfarlane}
parameter $q$ with $q=1$ giving bosons. Although we have as yet not made a
general analysis of such deformed algebras, we will consider the following
particular deformation of the usual harmonic oscillator algebra:
\begin{equation}
a_{q}a_{q}^{\dagger}-q^{-1}a_{q}^{\dagger}a_{q}=q^{N_{q}}\label{qdeform}%
\end{equation}
where the $q$-operators $a_{q},a_{q}^{\dagger}$ and $N_{q}$ act on a Hilbert
space with a\ denumerable orthonormal basis  $\left\{  \left\vert
n\right\rangle _{q},n=0,1,2,\ldots\right\}  $ as follows:%
\begin{equation}
a_{q}^{\dagger}\left\vert n\right\rangle _{q}   =\left[  n+1\right]
^{1/2}\left\vert n+1\right\rangle _{q},\nonumber\\
a_{q}\left\vert n\right\rangle _{q} =\left[  n\right]  ^{1/2}\left\vert
n-1\right\rangle _{q},\label{deformedHilbertspace}\\
N_{q}\left\vert n\right\rangle _{q} =n\left\vert n\right\rangle
_{q}\nonumber
\end{equation}
and
\begin{equation}
\left[  x\right]  \equiv\frac{q^{x}-q^{-x}}{q-q^{-1}}.
\end{equation}
Clearly when $q=1$ the boson case is recovered. We can consider the $q$
generalisation of $\mathcal{H}$.\ The resulting $q$ operator is
\begin{equation}
\mathcal{H}_{q}=\hbar\nu a_{q}^{\dag}a_{q}+\frac{1}{2}\hbar\Omega\sigma
_{3}^{\left(  1\right)  }+\frac{1}{2}\hbar\Omega\sigma_{3}^{\left(  2\right)
}+\hbar\gamma\sum_{i=1}^{2}\left(  a_{q}\sigma_{+}^{\left(  i\right)  }%
+a_{q}^{\dag}\sigma_{-}^{\left(  i\right)  }\right)  \label{qHamiltonian}%
\end{equation}
and there is still the natural $q$ analogue $\mathcal{E}_{n}^{\left(
q\right)  }$ of the subspace $\mathcal{E}_{n}$ which is invariant under
$\mathcal{H}_{q}$ and spanned by the basis
\begin{equation}
\left\vert e_{1}^{\left(  n\right)  }\right\rangle _{q}  \equiv\left\vert
\uparrow^{\left(  1\right)  },\,\uparrow^{\left(  2\right)  }\right\rangle
\left\vert n\right\rangle _{q},\\
\left\vert e_{2}^{\left(  n\right)  }\right\rangle _{q}   \equiv\left\vert
\uparrow^{\left(  1\right)  },\,\downarrow^{\left(  2\right)  }\right\rangle
\left\vert n+1\right\rangle _{q},\\
\left\vert e_{3}^{\left(  n\right)  }\right\rangle _{q} \equiv\left\vert
\downarrow^{\left(  1\right)  },\,\uparrow^{\left(  2\right)  }\right\rangle
\left\vert n+1\right\rangle _{q},
\end{equation}
and
\begin{equation}
\left\vert e_{4}^{\left(  n\right)  }\right\rangle _{q}\equiv\left\vert
\downarrow^{\left(  1\right)  },\,\downarrow^{\left(  2\right)  }\right\rangle
\left\vert n+2\right\rangle _{q}.
\end{equation}
A very similar calculation to the bosonic case yields
\begin{equation}
\mathcal{H}_{q}^{\left(  n\right)  }=\left(
\begin{array}
[c]{cccc}%
\Omega+\left[  n\right]  \nu & \gamma\sqrt{\left[  n+1\right]  } & \gamma
\sqrt{\left[  n+1\right]  } & 0\\
\gamma\sqrt{\left[  n+1\right]  } & \left[  n+1\right]  \nu & 0 & \gamma
\sqrt{\left[  n+2\right]  }\\
\gamma\sqrt{\left[  n+1\right]  } & 0 & \left[  n+1\right]  \nu & \gamma
\sqrt{\left[  n+2\right]  }\\
0 & \gamma\sqrt{\left[  n+2\right]  } & \gamma\sqrt{\left[  n+2\right]  } &
\left[  n+2\right]  \nu-\Omega
\end{array}
\right)  .\label{qdeformedinteraction}%
\end{equation}
Again the vector $\left\vert e_{2}^{\left(  n\right)  }\right\rangle
_{q}-\left\vert e_{3}^{\left(  n\right)  }\right\rangle _{q}$ is an
eigenvector with eigenvalue $\left[  n+1\right]  \nu$ . Hence this model using
generalised statistics based on the Jaynes-Cummings framework also does not
lead to the $\omega$-effect.

One cannot gainsay that other more complicated models of `thermal' baths may
display the $\omega$-effect but clearly a rather standard model rejects quite
emphatically the possibility of such an effect. This by itself is very
interesting. \ It shows that the $\omega$-effect is far from a generic
possibility for space-time foams. Just as it is remarkable that various
versions of the paradigmatic `thermal' bath cannot accommodate the $\omega
$-effect, it also remarkable that the D-particle foam model manages to do so
very simply.

\bigskip

\section{Conclusions}

{} In this work we have discussed two classes of space-time foam models, which
may characterise realistic situations of the (still elusive) theory of quantum
gravity. In one of the models, inspired by non-critical string theory, string
mattter on a brane world interacts with D particles in the bulk. Recoil of the
heavy D particles owing to interactions with the stringy matter produces a
gravitational distortion which has a backreaction on the stringy matter. This
distortion, consistent with a logarithmic conformal field theory algebra,
depens on the recoil velocity of the D-particle. It is modelled by a
stochastic metric and consequently can affect the different mass eigenstates
of neutral mesons.There are two contributions to the stochasticity. One is due
to the stochastic aspects of the recoil velocity found in low order loop
amplitudes of open or closed strings ( performed explicitly in the bosonic
string theory model) while the other is the leading behaviour of an infinite
resummation of the subleading order infrared divergences in loop perturbation
theory. The former is assumed to fluctuate randomly, with a dispersion which
is viewed as a phenomenological parameter. The latter for long enough capture
times of the D-particle with stringy matter gives a calculable dispersion
which dominates the phenomenological dispersion for the usual expectations of
the quantum gravity scale. This gives a prediction for the order of magnitude
of the CPTV $\omega$-like effect in the initial entangled state of two neutral
mesons in a meson factory based on stationary (non degenerate) perturbation
theory for the gravitational dressing of the correlated meson state.The
Klein-Gordon hamiltonian in the induced stochastic gravitiational field was
used for the calculation of the dressing.The order of magnitude of $\omega$
may not be far from the sensitivity of immediate future experimental
facilities, such as a possible upgrade of the DA$\Phi$NE detector or a B-meson factory.

This causes a CPTV $\omega$-like effect in the initial entangled state of two
neutral mesons in a meson factory, of the type conjectured in \cite{Bernabeu}.
Using stationary perturbation theory it was possible to give an order of
magnitude estimate of the effect: the latter is momentum dependent, and of an
order which may not be far from the sensitivity of immediate future
experimental facilities, such as a possible upgrade of the DA$\Phi$NE detector
or a B-meson factory. A similar effect, but with a sinusoidal time dependence,
and hence experimentally disentanglable from the initial-state effect , is
also generated in this model of foam by the evolution of the system.

A second model of space-time foam, that of a thermal bath of gravitational
degrees of freedom, is also considered in our work, which, however, does not
lead to the generation of an $\omega$-effect. Compared to the initial
treatment of thermal baths further significant generalisations have been
considered here. In the initial treatment (using a model derived in the
rotating wave approximation) the structure of the hamiltonian matrix was block
diagonal, the blocks being invariant 4-dimensional subspaces.The real
signifcance of the block diagonal structure was unclear for the absence of the
omega effect in this model and has remained a matter for debate. We have
analyzed the model ( now without the rotating wave approximation) and to \ our
surprise the omega effect is still absent. Furthermore we have also considered
$q$ oscillators for the heat bath modes since from Matrix theory we know that
energetic D-particles satisfy $q$ statistics. For the model of $q$ statistics
that we have adopted there is again no $\omega$-effect in the $q$ version of
the thermal bath.

It is interesting to continue the search for other (and more) realistic models
of quantum gravity, either in the context of string theory or in alternative
approaches, such as loop quantum gravity, exhibiting intrinsic CPT Violating
effects in sensitive matter probes. Detailed analyses of global data in
relation to CPTV, including very sensitive probes such as high energy
neutrinos, is a good way forward.

{ \ }

\section{Acknowledgements}

It is a pleasure to acknowledge partial support by the European Union through
the FP6 Marie Curie Research and Training Network UniverseNet (MRTN-CT-2006-035863).

\end{document}